*Article*

# Adaptive Measurement-Based Policy-Driven QoS Management with Fuzzy-Rule-based Resource Allocation

**Suleiman Y. Yerima** [1,*], **Gerard P. Parr** [2], **Sally I. McClean** [2] **and Philip J. Morrow** [2]

1 Centre for Secure Information Technologies, Queen's University, Belfast, Northern Ireland, BT3 9DT, UK
2 School of Computing and Information Engineering, University of Ulster, Northern Ireland, BT52 1SA, UK; E-Mails: gp.parr@ulster.ac.uk (G.P.P.); si.mcclean@ulster.ac.uk (S.I.M.); pj.morrow@ulster.ac.uk (P.J.M.)

**\*** Author to whom correspondence should be addressed; E-Mail: s.yerima@qub.ac.uk; Tel.: +44-028-9097-1896.

**Abstract:** Fixed and wireless networks are increasingly converging towards common connectivity with IP-based core networks. Providing effective end-to-end resource and QoS management in such complex heterogeneous converged network scenarios requires unified, adaptive and scalable solutions to integrate and co-ordinate diverse QoS mechanisms of different access technologies with IP-based QoS. Policy-Based Network Management (PBNM) is one approach that could be employed to address this challenge. Hence, a policy-based framework for end-to-end QoS management in converged networks, CNQF (Converged Networks QoS Management Framework) has been proposed within our project. In this paper, the CNQF architecture, a Java implementation of its prototype and experimental validation of key elements are discussed. We then present a fuzzy-based CNQF resource management approach and study the performance of our implementation with real traffic flows on an experimental testbed. The results demonstrate the efficacy of our resource-adaptive approach for practical PBNM systems.

**1. Introduction**

Next generation networks are increasingly being characterized by the convergence of heterogeneous fixed and wireless access network technologies such as WiMAX, WLAN, xDSL, and Passive Optical Networks, *etc.* towards interconnectivity with IP-based core networks. The management and control of network resources in such converged domains so as to provide the QoS that will in turn deliver satisfactory end user QoE poses significant technical challenges, given the complexity introduced by the integration of diverse QoS control mechanisms that may be deployed within each access technology on the end-to-end transport path. A promising network management approach that can be employed to overcome the challenges is the Policy-Based Network Management (PBNM) paradigm. PBNM allows for configuration and control of the network as a whole thus eliminating the need to individually manage each network entity. Thus, PBNM can simplify the administration of complex operational characteristics of a network, including QoS, access control, network security, and IP address allocation [1]. PBNM systems generally use hierarchical structures and will facilitate the management of next generation networks [2]. Moreover, PBNM provides the means for application transparency across existing and emerging access technologies which permit applications to be transport layer-agnostic when deployed [3]. It is also an effective means to equip networks with the ability to provide an automatic response to network conditions according to *pre-defined policies*, and is considered the best approach to ease the complex network configuration process involved in the integration of network services into a single large network [4,5]. These advantages of the PBNM approach and its well documented potential to facilitate QoS management in converged networks motivates its adoption as the underlying design paradigm for our proposed QoS management framework, CNQF (Converged Networks QoS Framework).

CNQF is a PBNM framework designed to provide homogeneous, unified, end-to-end QoS management over heterogeneous access technologies, together with scalable and adaptive resource control. From the operational point of view, CNQF comprises of distributed *Policy Decision Points* (PDPs) which handle various management and control decisions driven by *high-level declarative policies* and enforced at *Policy Enforcement Points* (PEPs) such as gateways, routers, access points, switches *etc.* within the converged network's management domain. CNQF will benefit operators through providing support for critical policy control and adaptation functionality needed to deliver a wide variety of high-value services with guaranteed QoS across fixed and wireless access technologies. Furthermore, functionalities for *context-aware* resource control have been incorporated within the CNQF architecture as a crucial PBNM enhancing feature to enable value-added capability to intelligently manage network resources and potentially increase revenue for network providers.

Our previous work presented the proposed CNQF architecture, detailing the subsystems, elements, and their interaction whilst also exemplifying some use case scenarios for *context-driven QoS control* [6]. In this paper we present the implementation of CNQF in Java, validating key elements functionality within a policy-driven QoS control scenario. Furthermore, the paper also presents an experimental evaluation of *fuzzy logic based resource-adaptive admission control for tiered services management* built into the CNQF framework prototype. Incorporating fuzzy logic into the PBNM framework potentially allows not only intelligent and adaptive resource control under dynamic operational conditions but also an intuitive and cost-effective means to integrate operator policies with extensible rule-based control.

The paper is organized as follows. Related work is reviewed next in Section 2, while Section 3 summarizes the CNQF framework architecture. Section 4 presents CNQF prototype development in Java and Section 5 explores the prototype validation by means of policy-driven QoS control of real traffic flows on a Linux based testbed. Section 6 presents the fuzzy-based CNQF resource management approach and its performance evaluation, while Section 7 concludes the paper.

## 2. Related Work

Standards bodies such as 3GPP, ETSI's TISPAN and IETF have defined a number of policy-based control architectures which can be found in [7–9] for example. TISPAN specified a Resource and Admission Control Subsystem (RACS) comprising of a Service-based Policy Decision Function (SPDF) and Access Resource Admission Control Function (A-RACF) [4,8]. Both SPDF and A-RACF interact with Policy Enforcement Points (PEPs) in the underlying transport network. Similarly, the policy framework in the 3GPP technical specifications defines a Policy Decision Function (PDF) as part of its policy-based architecture [7]. The 3GPP and TISPAN policy-based architectures share similarity with the IETF policy model which specified PEP and a (Policy Decision Point) PDP as part of its architecture [9]. Although the standards' policy frameworks define function blocks, interfaces and protocols to facilitate interoperability of the standards-compliant products from different vendors, no details are provided on how the various functionalities would be implemented. Our work aims to bridge this gap by not only leveraging similar policy-based approach to develop scalable solutions for QoS management and control of converged networks but also introduces important novel extensions such as context-awareness support in the design, and *fuzzy-rule-based* functionality in our implemented prototype in order to provide value-added enhancements to the basic PBNM operations.

With the continuous evolution of access technologies coupled with the need to address their interoperability and management requirements within converged architectures, PBNM systems are still actively being researched. PBNM systems have recently been studied in different contexts such as tactical networks [10], MPLS Networks [11], virtualization environments [12], Virtual Private Networks [13], IMS-enabled networks [2], *etc.* In [10], Kim *et al.*, present a policy-based QoS framework built on DiffServ and SNMP for the purpose of QoS control in *ad hoc* military environments. Although the authors demonstrated the efficacy of their framework using off-the-shelf technologies like we have undertaken in this paper, the design and application of their framework differs significantly from our CNQF approach. Moreover, unlike [10] our framework incorporates PBNM enhancements such as context management functionality and fuzzy-rule-based resource control as will be shown later in the paper. Similar to [10], Cha *et al.* in [14] and Carrilho and Ventura in [15] have also presented policy based management frameworks for DiffServ enabled networks. The same approach is extended to IMS networks by the authors of [2]. Whilst the system in [14] is built using EJB (Enterprise Java Beans), the framework in [15] relies on XML technologies. However, their work also did not consider the enhancements to the basic PBNM functionalities discussed, implemented and evaluated in this paper.

In [16], Sohail and Khanum investigate the use of fuzzy logic for network management systems. The article suggests that network management can benefit immensely from the use of fuzzy logic in various possible applications areas. They also present a fuzzy-based method for customized resource allocation decisions. In particular, a Fuzzy Bandwidth Broker (FBB) is proposed for distributed architecture management of Bandwidth Brokers in a DiffServ domain and also its performance optimization. Using a fuzzy rule base, various performance metrics are used to decide when to switch from centralized to distributed Bandwidth Broker architecture as well as the number and optimum placement of Bandwidth Broker entities. Different from [16], our work investigates the application of fuzzy logic to a different problem in network management *i.e.*, dynamic resource allocation for enhanced admission control especially for tiered services within practical PBNM systems, rather than Bandwidth Broker placement or their optimum topology selection.

Admission control is a crucial element of the PBNM-based management infrastructure as it provides the means to maintain QoS of the flows already present within the network. Several admission control schemes that have been previously studied are mainly classified as either *parameter-based* or *measurement-based* [17–19]. While parameter-based approaches may be easier to implement as they are based on estimates of parameters such as peak rate or effective bandwidth usage in the admission request of a new flow rather than actual network measurements, measurement-based approaches provide better network utilization in the presence of bursty traffic patterns as shown in [17]. In [19], the authors focus on evaluating the effectiveness of measurement-based admission control in 3G and 4G wireless networks, where the results of their study highlight the benefits of measurement-based admission control within wireless IP networks. Earlier work [20], proposes a time-window based measurement based admission control scheme for dynamic call admission control in ATM networks using marginal distributions of cell arrivals for aggregate bandwidth estimates. Similarly, [21–23], have also considered time-window based measurement approaches while [24] offered an implementation-based comparison of various estimators employed in measurement-based admission control schemes and [25] focused on measurement-based scheme with aggregate traffic envelopes. The authors of [26] proposed and evaluated an online measurement-based admission control scheme applied to variable bit-rate video where effective bandwidth is estimated using statistical parameters of the aggregate video traffic.

The aforementioned previous works have proposed several effective measurement based methods for admission control in single-class scenarios, but further studies are required to investigate their applicability, scalability and performance to multi-class scenarios in order to support tiered services. Also, given the requirement to rapidly deploy and support dynamic heterogeneous services within today's converging networks, practical PBNM systems should have scalable, extensible and cost-effective multi-class admission control solutions. Hence, in this paper, we considered a fuzzy logic based approach as it provides highly efficient control without requiring crisp measurements which must be available for traditional measurement-based approaches. Not only are monitoring overhead requirements for PBNM systems reduced with a fuzzy-based approach, but also the admission control process is better able to cope with imprecise or incomplete data compared to traditional measurement-based methods. Furthermore, a fuzzy-based approach is more flexible and easily extensible with multiple classes and therefore well suited to tiered services management than traditional measurement-based approaches.

Fuzzy logic has also been previously applied to admission control problems in wireless/cellular networks [27–32]. For instance, in [27], Huang, Chuang and Yang applied fuzzy logic to adaptive call admission control and bandwidth reservation to alleviate forced termination probability of multimedia handoffs in cellular systems. The authors showed the feasibility of their approach by means of computer simulations which indicated that bandwidth reservation using fuzzy logic system tuned by Particle Swarm Optimization has the potential to improve the admission of new calls whilst reducing forced termination of ongoing calls during handoff. Other papers such as [33,34] focus mainly on employing fuzzy logic for alleviating wireless/mobile network handoff problems.

Although fuzzy logic is becoming more popular for alleviating network resource management problems, to the best of our knowledge we are not aware of existing work in the literature that have experimentally studied its application in a real implementation of a PBNM system. Hence, this paper aims to explore enhancing PBNM operations by incorporating fuzzy-rule-based systems and to evaluate the implemented system within our CNQF PBNM prototype using real traffic. The next section describes the CNQF PBNM architectural framework and constituent elements.

## 3. CNQF Architecture

The authors of [1] emphasize the distinction between *QoS mechanisms* which support service differentiation (that determine *how* to achieve performance objectives for heterogeneous flows) and *QoS management and control* (concerned with *who* should be entitled to preferential treatment from the network). CNQF mainly addresses the latter (QoS management/control) through its framework architecture which can be adapted to support heterogeneous underlying *QoS mechanisms* (such as DiffServ, MPLS, WiMAX QoS, UMTS QoS *etc.*) that may be deployed in the access and/or core segments of the transport layer. The logical subsystems that make up the CNQF framework include: *Resource Management Subsystems (RMS), Measurement and Monitoring Subsystem (MMS) and Context Management and Adaptation Subsystems (CAS).* As shown in Figure 1, CNQF has distributed functional entities whose instances coordinate the resources of the transport network to enable closed-loop, scalable, end-to-end QoS control and resource management in converged networks (by leveraging existing QoS mechanisms in the transport plane).

*3.1. Resource Management Subsystem (RMS)*

The two key elements that make up the RMS are the Resource Brokers (RBs) and the Resource Controllers (RCs). The distributed instances of resource brokers which could be Wireless Access (WARB), Fixed Access (FARB) or Core Network (CNRB), perform policy-based resource management roles depending on the segment of the converged transport plane the RB oversees (see Figure 1). For instance, in a wireless access network, a WARB will administer admission control and bandwidth management policies that affect end-users present in that access domain. CNRB is responsible for resource management in a core network but would also interface with various WARBs and FARBs (if present) to allow interchange of resource availability status thus enabling end-to-end (global) resource management in the converged network. The RBs act as PDPs and communicate with one or more RCs which are located at various PEPs (*i.e.*, routers and gateways). The policy decisions in the RBs invoke *policy actions* which translate to a set of configuration and control functions undertaken by the RCs. In our previous work we demonstrated CNQF-based open loop configuration of routers for DiffServ operation with the RC functionality under RB command [35]. The policy-driven fuzzy-based resource control strategy described and evaluated in Section 6 of this paper is implemented within the Resource Broker as part of the RMS functionality (for controlling the edge network devices where admission control is administered).

**Figure 1.** Converged Networks QoS Management Framework (CNQF) subsystems and operational entities in a converged network scenario.

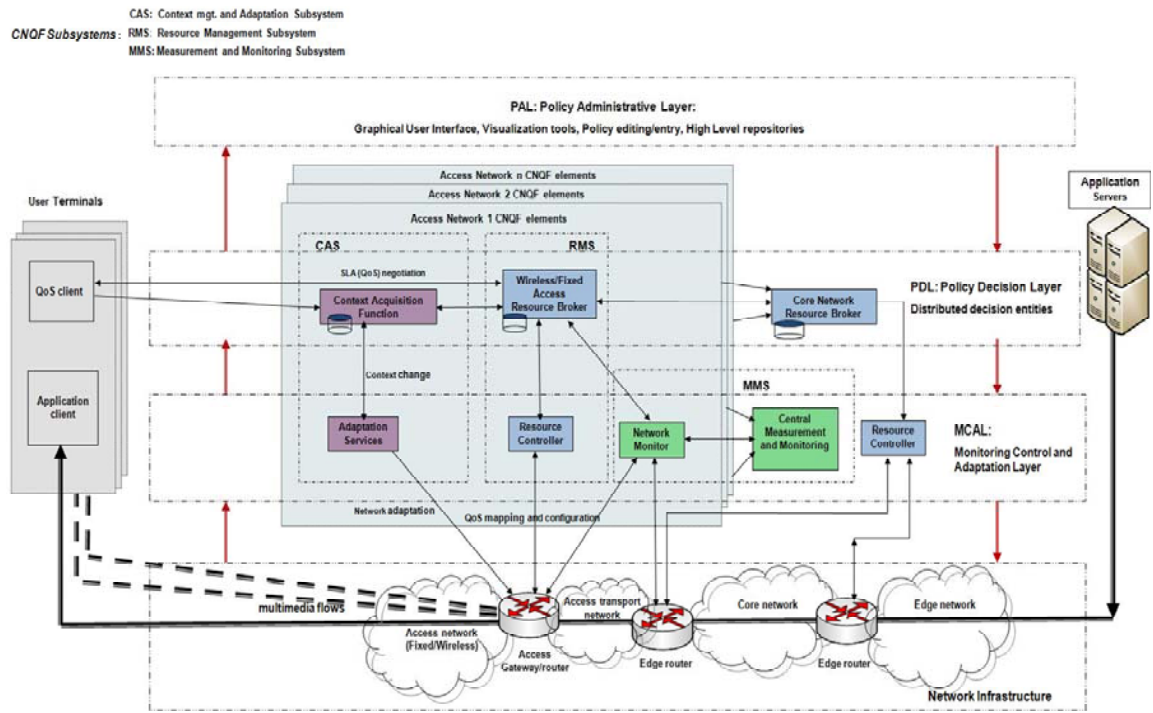

*3.2. Measurement and Monitoring Subsystem*

The MMS enables *closed-loop, adaptive, measurement-based control* within the CNQF architecture. MMS consists of distributed Network Monitors (NMs) which interface with the Policy Enforcement Points to provide *active* and *passive* measurement capabilities. In order to mitigate potentially excessive monitoring data interchange across the network, each NM is designed to collect and process measurement data locally and can then be invoked via a central measurement and monitoring element (CMM), to provide averages/summaries or asynchronous instantaneous measurement data. CMM could provide high level summaries that may be used to gauge network-wide health status through visualization interfaces in a centralized management station. MMS basically provides measurement and monitoring services to the other CNQF subsystems.

*3.3. Context Management and Adaptation Subsystem*

As shown in Figure 1, CNQF provides context-aware functionality through its Context Management and Adaptation Subsystem (CAS). CAS consists of distributed Context Acquisition Function blocks (CAF) instantiated in each access network. The CAFs are PDPs that execute context-aware or context-driven policies within the CNQF system. Each CAF elements has associated Adaptation Servers (ADs) which are function blocks that configure/reconfigure PEPs directly affected by context-driven policy decisions in the CAF. See [35] for further discussion on context-aware QoS control with CAS.

*3.4. CNQF Hierarchical Structure*

While functional elements within the subsystems are designed to be integrated horizontally, they also form part of a hierarchical structure as is typical of PBNM systems [2]. The hierarchical structure is also depicted in Figure 1. The top level consists of tools that provide centralized administrative capabilities such as Graphical User Interfaces (GUI) for interactive high level configuration, and policy entry/editing; visualization tools for network-wide configuration and status monitoring; and central repositories. The PDL (Policy Decision Layer) comprises of various Policy Decision Points (PDPs) *i.e.*, Resource Brokers of the RMS; these are centrally managed via the PAL (Policy Administrative Layer). The bottom layer is the MCAL (Measurement, Control and Adaptation) layer that consists of elements that directly interface with the Policy Enforcement Points (PEPs) in the transport network. The MCAL elements include the various RMS Resource Controllers, the ADs and the various MMS NMs.

## 4. CNQF Prototype and Testbed Implementation

Based on the proposed CNQF architecture described in the previous section, a working prototype has been implemented in Java. Presently, distributed entities in the RMS and MMS subsystems which provide capability for adaptive policy-based QoS management based on CNQF architecture have been built using the *NetBeans IDE 6.9 Java SDK*. Furthermore, a testbed has been implemented using Linux machines to evaluate the CNQF prototype. Brief descriptions are outlined in this section.

*4.1. CNQF Prototype Development*

4.1.1. RMS Implementation

A resource broker (RB) has been built where resource management *decision logic* are implemented. High level policies entered into the GUI policy editor are mapped into a set of *commands* within the decision logic in the RB (the logic implementation will be different for CN-, WA-, and FARBs functionalities). These commands are Java code segments that invoke the services of instances of other CNQF entities (such as network monitors and resource controllers) that are installed and running at the PEPs in the network. For instance, a high level CNQF policy that has an action part*: Configure Edge Router* will be mapped to the following RB Java code:

*new ResCon*
*ResCon.ConfigureEdge()*

This creates an instance of the Resource Controller class and calls its *ConfigureEdge()* method that provides edge router configuration services for the RB policies. The RC instances are installed on the PEPs (Linux-based routers on our testbed). The installed RC opens an RMI port (Remote Method Invocation, TCP port 1099) to listen for incoming RB commands. RC is implemented as a *ResConImpl* class. *ResConImpl* implements methods such as *ConfigureEdge(), ConfigureCore()* and a host of others that can be invoked by the RB via the RMI communication interface. Details of the methods' implementation will depend on the PEP type *i.e.*, whether it is a router, gateway or an access point and also the specific APIs available for interacting with the internal mechanisms. Thus, the *ResConImpl* class is the *wrapper class* which can be customized to wrap the functionality of specific QoS mechanisms within the heterogeneous PEPs thus *exposing a homogenous interface that enables technology independent Resource Brokers to be achieved*.

4.1.2. MMS Implementation

In order to enable closed-loop QoS control within CNQF prototype, the MMS network monitoring element has been implemented as a *netmon* class. The main feature of netmon class is the provision of functions to enable SNMP based measurements using available Management Information Blocks (MIBs). Thus CNQF can create and install instances of *netmon* at various PEPs and communicate with SNMP agents at the PEPs via the *netmon* instance. In order to monitor attributes such as bandwidth on a router interface, the *netmon* instance is supplied with the IP address and MIB OIDs (Object Id) representing the required attributes as well as the SNMP *community string*. The CNQF *netmon* class incorporates an SNMP agent which is built using *SNMP4j* [36], an open source object oriented SNMP API for Java managers and agents. SNMP4j supports command generation (manager mode) and command responding (agent mode) as well as synchronous and asynchronous requests. Table 1 depicts the RFC 1213 MIB OIDs used within the *netmon* class for bandwidth monitoring. From the MIB attributes, *netmon* calculates the bandwidth using:

$$\text{BW (bits/s)} = ((O(t) - O(t - \Delta t))*8)/\Delta t \tag{1}$$

where $\Delta t$ is the interval between two SMNP *get* operations which are used to read the MIB values $O(t)$ which is basically a counter indicating the number of octets sent (ifOutOctets) or received (IfInOctets) on the network interface. Since the MIB variables are stored as counters, two poll cycles are taken by the *netmon* instance and the difference is calculated to obtain the bandwidth. Utilization is calculated using:

$$\text{BWU (\%)} = ((O(t) - O(t - \Delta t))*8*100)/(\Delta t * \text{ifSpeed}) \tag{2}$$

**Table 1.** MIB OIDs used in CNQF *netmon* class for bandwidth monitoring (RFC 1213).

| MIB Object | Description | OID |
|---|---|---|
| ifInOctets | The total number of octets received on the interface, including framing characters | 1.3.6.1.2.1.2.2.1.10.2 |
| ifOutOctets | The total number of octets transmitted out of the interface, including framing characters | 1.3.6.1.2.1.2.2.1.16.1 |
| ifSpeed | An estimate of the interface's current bandwidth in bits/sec | 1.3.6.1.2.1.2.2.1.5.1 |

*4.2. Testbed Implementation*

The testbed used for development of CNQF and validation of its functionalities is implemented using Linux systems. The configuration is shown in Figure 2. The testbed consists of two Linux-based edge routers and two Linux-based core routers. These elements constitute the PEPs each having an instance of CNQF RC (ResCon) that interacts with the Linux router kernel to set various parameters that enable configuration/reconfiguration of QoS provisioning to suit the high-level policies processed by the RB. The Linux TC (traffic control) utility in the kernel provides commands for implementing packet marking, classification, queuing disciplines, and policing of flows (thus enabling the transport plane mechanisms required to provide QoS and resource control). In the testbed developed, the RCs employ TC commands for low level configuration which have equivalent mappings to RB Java code that implement the high level *policy actions*. The testbed elements include:

- *CNQF management station*: houses central CNQF management application with the GUI policy editing tool and RMS CNRB implemented in Java which invokes policy actions via remote RCs (ResCon instances) installed at the PEPs (edge and core routers in the testbed).
- *Edge routers*: Ubuntu 10.0.4 Linux PC with 2.66 GHz Intel Xeon, 3GB RAM, configured as edge routers with TC utility installed to enable configuration of the router interface(s) for ingress packet marking, and for egress classification, queuing and policing via RC's response to CNQF policy decisions.
- *Core routers*: Ubuntu 10.0.4 Linux PCs with 2.66 GHz Intel Xeon, 3GB RAM, with TC utility installed to enable configuration of packet classifiers and filters through CNQF policies also via an RC.
- *Traffic generators*: The *ntools* traffic generator is used to generate multi-client traffic with different flow characteristics including constant bit rate (CBR), On-Off traffic, and variable bit rate (VBR) traffic. *ntools* is an open source versatile traffic generator, analyzer and network emulator package for Linux. The multi-client generation capability with packet field injection is used to emulate traffic from multiple access networks.

**Figure 2.** Linux-based testbed for CNQF development and experimental evaluation.

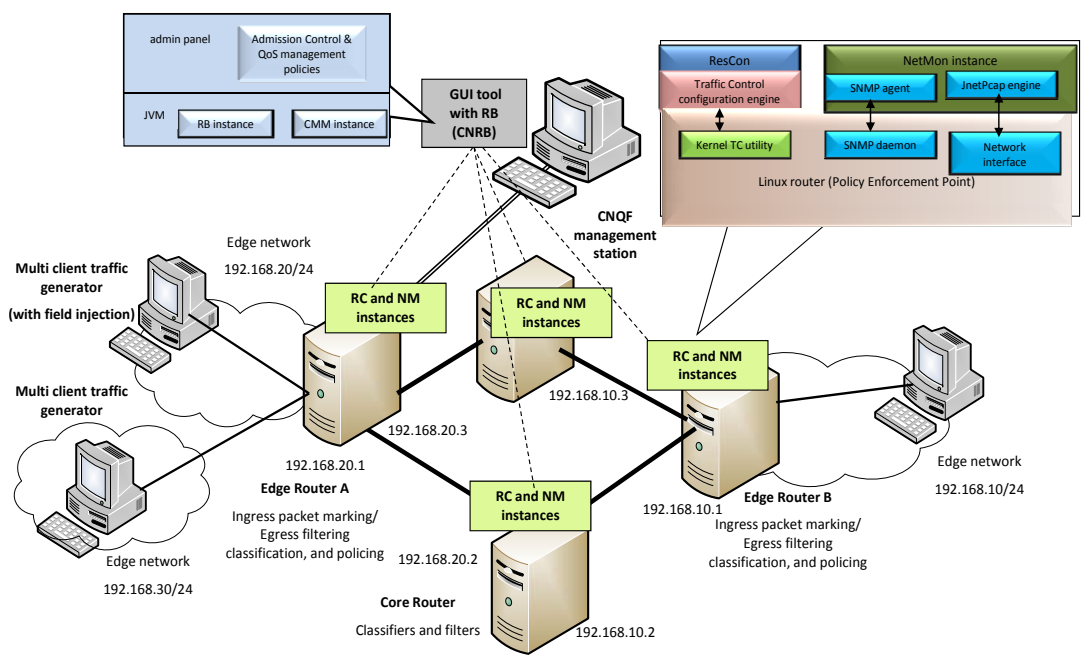

## 5. Policy-based Adaptive QoS Management: CNQF Runtime Validation

The elements within the CNQF subsystems described in Section 3 interoperate under the supervision of a central policy-based CNQF management application to co-ordinate the transport plane resources. The distributed entities exchange messages in response to events that trigger policy decisions which facilitate adaptive QoS management strategies. The network monitors allows for a closed-loop system enabling adaptive QoS configuration policies and other resource management functionalities such admission control. As static QoS configurations may not always meet service performance requirements in converged networks due to highly dynamic network conditions, *adaptive policy-based* control supported by the CNQF architecture can provide efficient network resource utilization and maintain service flow QoS requirements.

*5.1. Initial Configuration Policies*

For a QoS domain under CNQF administration, the initial configuration policies are entered into the GUI policy tool. These policies govern the configuration of the PEPs (routers) in the transport network with specific QoS provisioning capabilities. Current CNQF implementation supports DiffServ QoS mechanisms; initial/adaptive runtime management policies that utilize DiffServ QoS mechanisms can therefore be specified by an administrator. Extension of CNQF prototype to support others QoS technologies is such as MPLS can be achieved without rebuilding the PAL, PDL and most of the MCAL elements. The only requirement is the RC extension (using the wrapper functionality of the *ResConImpl* class); this is one major advantage of the CNQF architecture.

The CNQF GUI has provision for the DiffServ Code Points (DSCP) codes necessary for marking packets for service differentiation (in the edge routers) to be specified along with the selection of queuing discipline for implementing the per-hop-behaviors (PHB) for the QoS domain. Once these are entered by the admin, the relevant RB instance is called to process the information and will in turn invoke the associated RC instances to configure the routers with the selected configuration parameters. Once the PEPs have been configured, the RC instance returns a status message displayed in the CNQF GUI's status pane. Next, the policies relating to how specific flows, flow types or group of flows identified by *identifier* are to be handled (initial static QoS provisioning) are entered in the GUI. The format used is:

*If (flow==identifier) then (policy action)*

Here, the *identifier* can be a flow's IP address or a range of IP addresses (for group of flows) or other unique flow identifiers. This will normally be used for configuring flows for specific preferential QoS treatment (e.g., for Expedited Forwarding PHB in DiffServ domain). If not configured, the default best effort treatment is applied to a flow. Each of these policy rules adds an entry to a look-up table specifying which flows have which DSCP and therefore a corresponding PHB. Through the use of the look-up table each flow's current QoS configuration state is maintained. The RC employs the look-up table to add or modify Linux TC filter parameters within a file which the kernel employs for configuration of the router to implement the required DiffServ PHB.

*5.2. Runtime Adaptive Policies*

CNQF adaptive policies are triggered by temporal events such as congestion in the network. With the run time adaptive policies, the RB can invoke reconfigurations of QoS management via the associated RC instance, but will require monitoring reports which are provided by the netmon instance running on the PEP (routers). The message sequence diagram of the procedure is shown in Figure 3. In this particular test scenario, when bandwidth utilization on a link is high (indicative of congestion situation), the PEPs are reconfigured via a runtime policy to mark a specific flow with Expedited Forwarding DSCP (0x2e). This run time policy is specified in the GUI policy tool as follows:

- IF (*bandwith_utilization*)==high THEN (mark *flow_id* with DSCP = 0x2e).

**Figure 3.** CNQF enabled adaptive measurement-based QoS management procedures scenario.

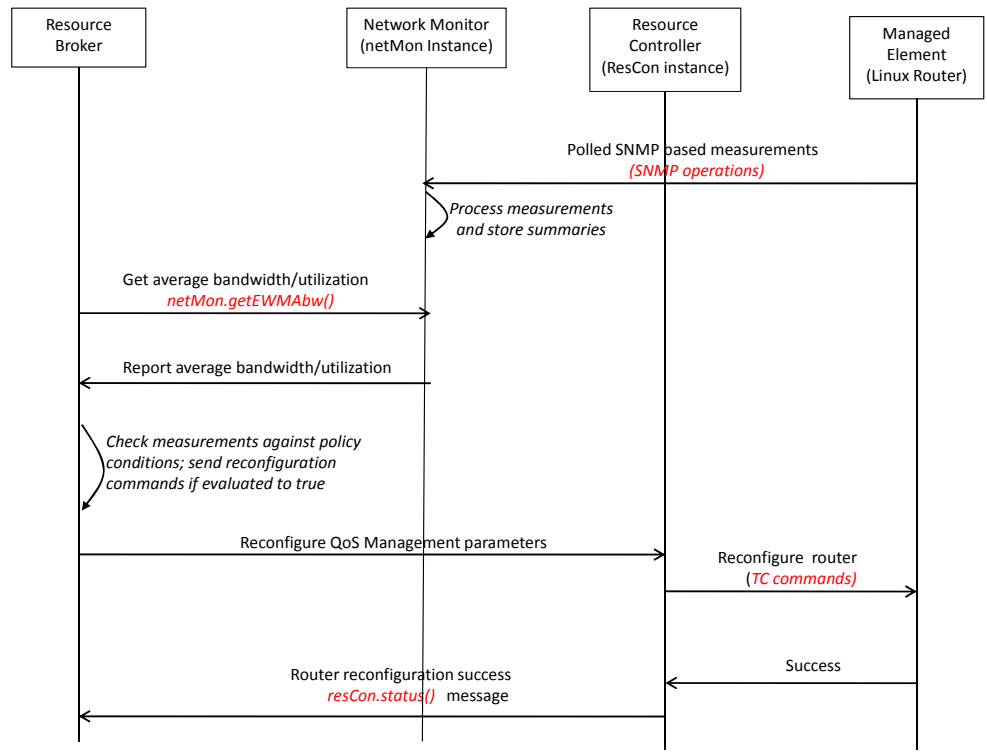

The runtime policies for CNQF measurement-based adaptive QoS configuration are specified in the following format:

- IF (*temporal_event*) == *state* THEN (*action*)

Here, *temporal_event* is a monitored event which may be mapped to specific *netmon* measurement ranges; while predefined RB-compliant commands invoke specific set of procedures in the RC instances to implement the required policy *action* when the condition *state* is fulfilled. At the same time the look-up table is updated with the changed flow's QoS specification. A temporal_event such as bandwidth utilization is monitored by the *netmon* instance in the PEP and the state is reported back to the RB via RMI protocol when a trigger threshold is exceeded. RB can also poll the *netmon* instance using the *netmon.getEWMA()* RMI call to determine whether the policy condition is satisfied upon which the policy action (reconfiguration) commands are sent to the RC. The RC then returns the status of the configuration action as success or failure which can be seen in the GUI status pane.

*5.3. Validation Tests*

In order to validate the CNQF prototype for adaptive policy-based QoS management we performed some tests on the Linux-based testbed depicted in Figure 2 where the prototype was deployed. High level policies were entered into the policy editing tool and the effects on the underlying network traffic were observed. The behavior of selected traffic flows under observation indicated whether the adaptive policies were successfully applied or not. First, the initial static configuration policies were applied via the GUI policy tool. This set of policies was intended to configure the network initially with DiffServ capabilities. Next, another set of policies for runtime dynamic reconfiguration to govern QoS control of selected flows in response to network measurements were entered. These allowed us to validate the CNQF prototype capability for DiffServ configuration as well as adaptive QoS management capability of the implementation.

5.3.1. Initial Configuration Validation

In this test, the DiffServ configuration parameters were entered via the CNQF policy editing tool and when the RB was invoked it successfully processed the request and configured the edge routers and the core routers as expected. These were indicated by the '*success*' status message displayed on the GUI's status pane which were returned by the remote RC instances running on the Linux routers. Note that this configuration must be successful before any adaptive runtime QoS management policies can work on the testbed. This is because the runtime policies rely on the correct functioning of the DiffServ mechanisms for QoS control.

5.3.2. Runtime Adaptive QoS Control Validation

In order to validate this aspect, a policy stipulating the reconfiguration of QoS parameters in the routers so as to provide Expedited Forwarding DiffServ treatment to a given CBR test flow was entered into the CNQF GUI tool. The condition stipulated was that bandwidth utilization state of 80% on the ingress edge router outbound interface should be exceeded (indicating high traffic load). Without an adaptive CNQF policy, increasing traffic load from other flows will starve the test flow of resources leading to rapid QoS degradation. In order to alleviate this, the policy enabled preferential QoS treatment to be applied to the test flow at all nodes in the DiffServ domain so that its QoS requirements can be maintained.

Figure 4 illustrates increasing traffic load generated from the *ntools* traffic generator which injects new flows into the DiffServ domain with Poisson arrivals and exponentially distributed average lifetime of 120 s. As seen from the corresponding measurement in Figure 5, a 1Mbit/s CBR test flow traversing the same DiffServ domain is initially unaffected when the traffic load in the network is low, and a slight degradation in bit rate is noticeable as the aggregate traffic increases. When the traffic load causes the bandwidth utilization threshold given in the policy condition to be exceeded, the action part of the adaptive policy is invoked consequently marking the packets to be treated with Expedited

Forwarding PHB thus restoring the test flow's QoS through priority queuing and scheduling. It can be seen that the reconfiguration takes place at time 70 s because the packet measurements filtered on DSCP == 0x2e (EF DSCP) from this time onwards coincides with the test flow measurement filtered on its IP address. The same test was repeated with the background flows applied but without enabling an adaptive QoS management policy for the test flow. Figure 6 shows the bit rate of the test flow degrading considerably due to the absence of adaptive runtime policy as the background traffic increases due to the absence of adaptive runtime policy control.

**Figure 4.** Increasing traffic flow injected into the DiffServ-enabled testbed.

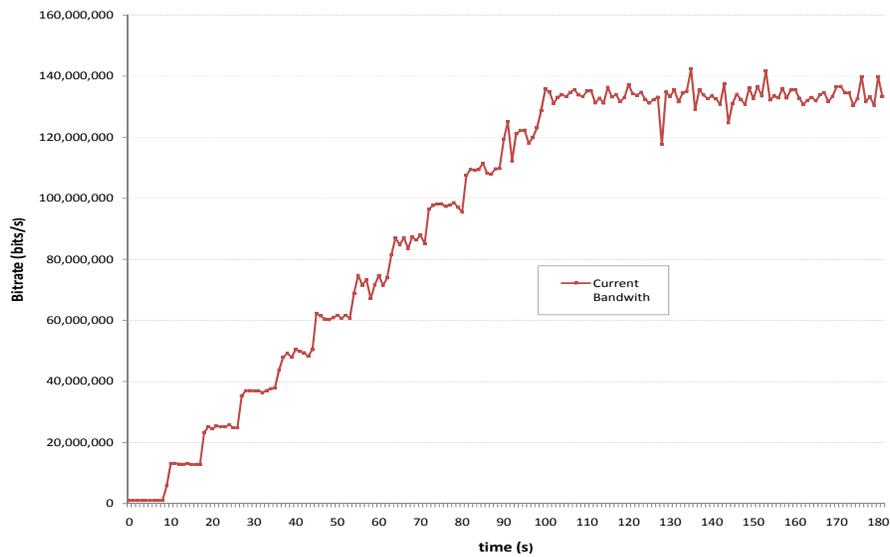

**Figure 5.** Effect of adaptive policy QoS reconfiguration on test flow with increasing background traffic load.

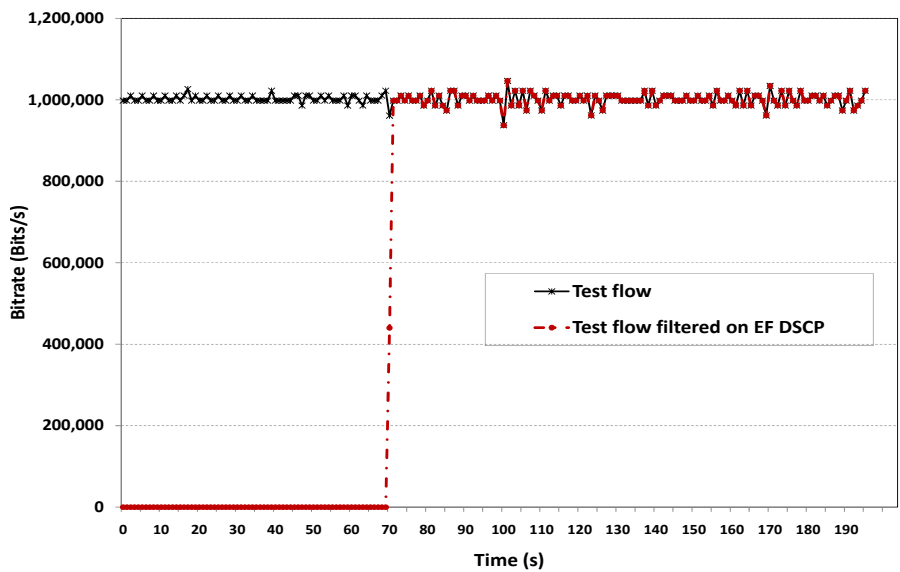

**Figure 6.** Effect of increasing background traffic load on test flow without adaptive CNQF reconfiguration policy.

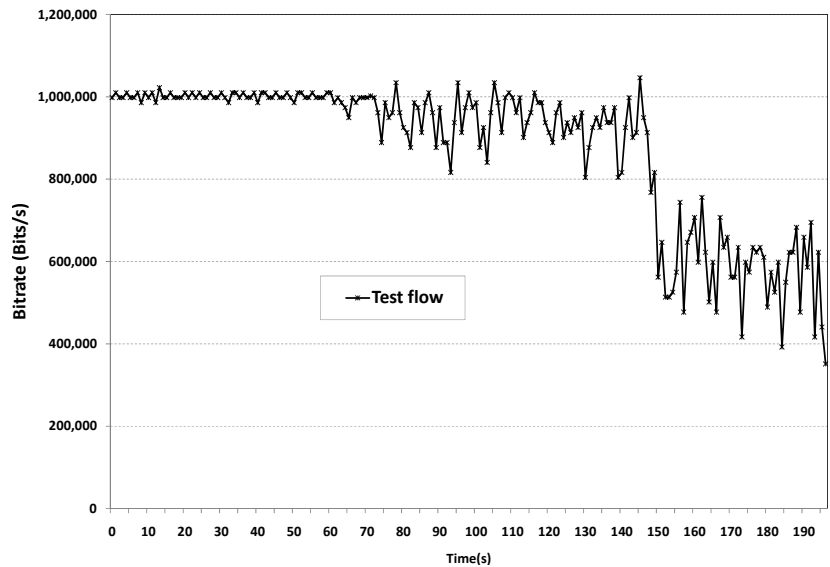

The test results illustrate the effect of the QoS control policies thus validating the operation of the CNQF prototype implementation and the experiments verified the interaction of the various subsystem elements built. Next, we explore *adaptive resource allocation* for enhanced admission control using implementation of fuzzy rule base system within the CNQF resource broker. As mentioned earlier, the motivation for implementing a fuzzy system within the RB is to enable effective and extensible means of coupling operational policies with rule-based control of network resources in order to achieve more efficient resource utilization as well as flexible tiered service provisioning.

## 6. Fuzzy Based Adaptive Resource Allocation

As shown in [16], fuzzy logic can be integrated with management tasks to optimize and customize their performance. This section presents the implementation of a fuzzy rule based system within the CNQF RB and *its use case scenario for adaptive class based resource allocation to support policy-driven admission control*. The ability to provide class based differentiated services (or service tiers) is one of the characteristics of QoS-enabled networks. Thus the mechanism for allocation of resources per service class has a critical impact on the availability of the services and efficient utilization of network resources. Because fuzzy logic allows approximate reasoning without requiring crisp and definite values (unlike traditional measurement-based methods), it can produce intelligent results for management purposes with minimal monitoring overhead. For this reason, fuzzy logic based approaches are desirable within evolving management systems as control and management tasks increase in complexity placing more demand on monitoring of resources, services and network elements.

*6.1. Fuzzy Logic System Implementation in CNQF*

A Fuzzy Rule Based (FRB) system is developed within the CNQF RMS Java prototype to provide *decision support* in the admission control process through adaptive (bandwidth) resource allocation between the existing service classes. The capability to handle uncertainty and imprecision as network and traffic conditions experience sudden, rapid and unexpected variation makes the fuzzy approach promising within dynamic operational environments. Thus fuzzy logic is proposed as a crucial component of the admission control system of CNQF RMS as shown in Figure 7.

**Figure 7.** FRB component implemented within CNQF Resource Management Subsystem.

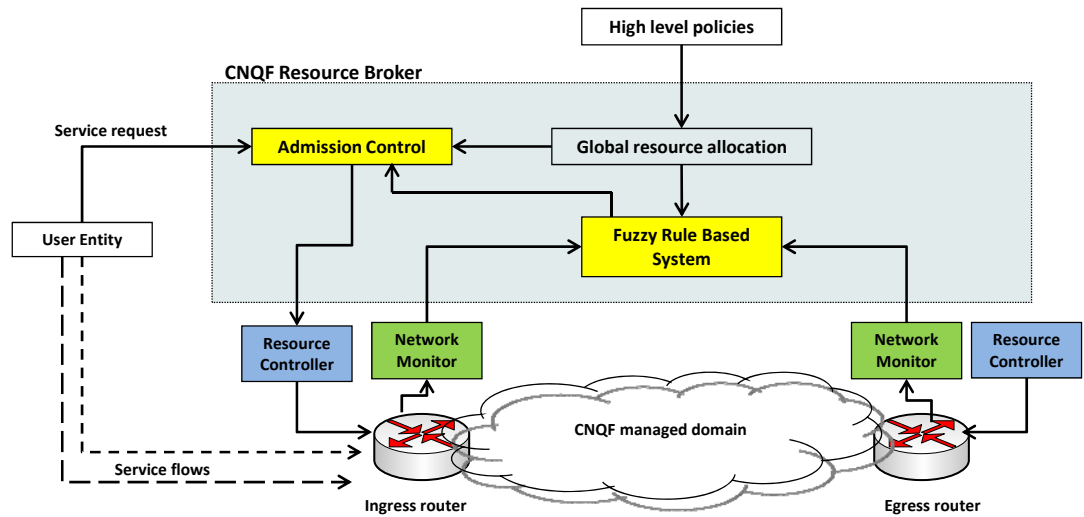

The FRB component is designed as a mechanism for dynamically controlling the global resource allocation parameters for the traffic classes with temporal changes in the network and traffic conditions. The FRB component fine-tunes the policy-based resource allocation process within the RB admission control entity thus making it *adaptive* and *resource-aware*. The motivation is to combine high level policies with resource-aware rule-based control so as to exploit the flexibility of fuzzy logic and improve system performance. Furthermore, the incorporation of fuzzy logic allows for easier extension of functionality without rebuilding the system from scratch; whereas this is difficult or impossible with traditional measurement-based admission control. This leads to cost-effective, easier-to-modify admission control and resource allocation within the CNQF management framework. For example, the rules governing the allocation can be altered or the number of classes changed without re-implementing the RB algorithms or the underlying control and configuration mechanisms in the PEPs.

*6.2. FRB Enhanced Class-based Policy-Driven Resource Control*

Let the service classes in the CNQF managed domain be denoted by $C_1$, $C_2$ and $C_3$ with corresponding per class bandwidth allocations $B_1$, $B_2$ and $B_3$ respectively from a maximum available bandwidth $B_T$. $B_1$, $B_2$ and $B_3$ are parameters stipulated within the high level policies in the policy management tool and can be altered to suit operator policies without affecting the operation of the underlying CNQF RMS admission control functionality. A dynamic resource allocation strategy is implemented within the RB whereby:

$$B_{th}^j = \{\max(B_j, B_{FLS}^j)\} | \forall\, j \in [1,2,3] \tag{3}$$

$B_{FLS}$ represents the output of the fuzzy logic system which varies with the traffic load per class within the network and $B_{th}^j$ is current bandwidth allocation for the $j^{th}$ service class.

The fuzzy logic system, $F$ is characterized by a membership function $\mu_F(x)$ where $x \in X$. The system is expressed as: $F = \{(x, \mu_F(x)) | \forall x \in X\}$ where the membership grade of $x$, $\mu_F(x)$, is a crisp value in [0, 1]. The fuzzy logic system is represented by fuzzy IF-THEN rules. These rules map the *input space* to the *output space* in a linguistic manner that can also allow the designer to incorporate initial knowledge of the operational domain, as well as depict intuitively the relationship between the input and output variables within the system. The $z^{th}$ rule of the fuzzy logic system having $q$ inputs $x_1 \in X_1, \ldots, x_q \in X_q$ and one output $y \in Y$ is expressed as:

*Rule*: IF $x_1$ is $F_1^z$ and/or $x_2$ is $F_2^z$ and/or…and/or $x_p$ is $F_q^z$ THEN $y^z = C^z$ (4)

where $F_i^z$ ($i = 1, 2, \ldots, q$) are *fuzzy antecedent sets* represented by their membership functions $\mu_{F_i}^z$, and $C^z$ is a fuzzy *consequent set* represented by the membership function $\mu_C^z$.

The output is obtained by combining the outputs from the $M$ rules in the following form:

$$y(x) = \frac{\int_M T_{k=1}^q \mu_{F_k}(x_k) y \, dy}{\int_M T_{k=1}^q \mu_{F_k}(x_k) \, dy} \quad (5)$$

where $T$ is a *t-norm* operation (*i.e.*, min or max operation) and $x$ is the input vector. Finally, the centroid defuzzification method is then used to compute from (5), a corresponding crisp output for the given input vector.

The FRB system implemented within the CNQF RB has the following characteristics. The inputs to the fuzzy system are the Class 1, Class 2 and Class 3 network loads. Class 1 network load is expressed by $\min\{b_1/B_T, B_1/B_T\}$ where $b_1$ is the current measured aggregate bandwidth consumption by Class 1 flows, while $B_1$ is set by the high level policy as mentioned earlier. $B_T$ is the total available bandwidth for all traffic. The input is modeled by three fuzzy sets with the universe of discourse given by $[0, B_1/B_T]$:

(1) *Light Load (LL)*, modeled by a Gaussian membership function with parameters mean $\beta = 0$, and variance $\sigma = (0.17 \, B_1/B_T)$.
(2) *Medium Load (ML)*, modeled by a Gaussian membership function with parameters mean $\beta = (0.5 \, B_1/B_T)$ and $\sigma = (0.13 \, B_1/B_T)$.
(3) *Heavy Load (HL)*, also modeled by a Gaussian membership function with parameters $\beta = B_1$, and $\sigma = (0.17 \, B_1/B_T)$.

Similarly, the Class 2 and 3 input variables expressed by $\min\{b_2/B_T, B_2/B_T\}$ and $\min\{b_3/B_T, B_3/B_T\}$ are *fuzzified* with identical membership functions and parameters as Class 1 with universe of discourse given by $[0, B_2/B_T]$ and $[0, B_3/B_T]$ respectively.

There are three output variables from the FRB, each indicating the current level of allocated bandwidth per class. The variables are the *Class1Res*, *Class2Res*, and *Class3Res* which denote the level of resource allocation per class. The variables have identical fuzzy sets as follows:

(1) *Light Allocation (LA),* modeled by a Gaussian membership function with parameters mean $\beta = 0$, and variance $\sigma = 0.17$.
(2) *Medium Allocation (MA),* modeled by a Gaussian membership function with parameters mean $\beta = 0.5$, and variance $\sigma = 0.13$.
(3) *High Allocation (HA),* also modeled by a Gaussian membership function with parameters mean $\beta = 1$, and variance $\sigma = 0.17$.

The Gaussian membership function is given by:

$$f(x) = \exp\left[\frac{-0.5(x-\beta)^2}{\sigma^2}\right] \quad (6)$$

where $\beta$ is the mean and $\sigma$ is the variance. The input membership function plots for Class 1 with $B_1 = 0.3B_T$, and Class2 with $B_2 = 0.4B_T$, are shown in Figures 8 and 9 respectively. Note that with the same universe of discourse as Class 1, Class 3 has an identical input membership function plot as Class 1. The corresponding resource allocation surface maps for Class 1 with Class 2 and Class 3 load variation is shown in Figure 10 while the resulting surface plot for Class 2 and Class 1 load variation is depicted in Figure 11. These parameters are used in the experiments discussed in the sub-sections that follow.

**Figure 8.** Input membership function plots for Class 1 with $B_1 = 0.3B_T$.

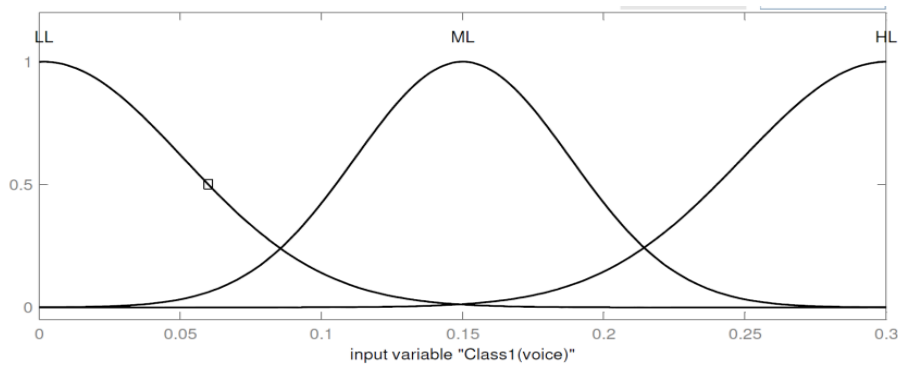

**Figure 9.** Input membership function plots for Class 2 with $B_2 = 0.4B_T$.

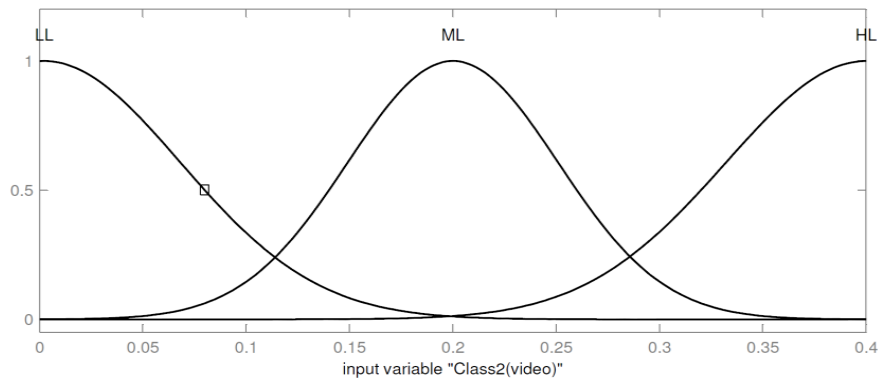

**Figure 10.** Sample output Resource allocation surface map for Class 1 with varying Class 2 and Class 3 network loads.

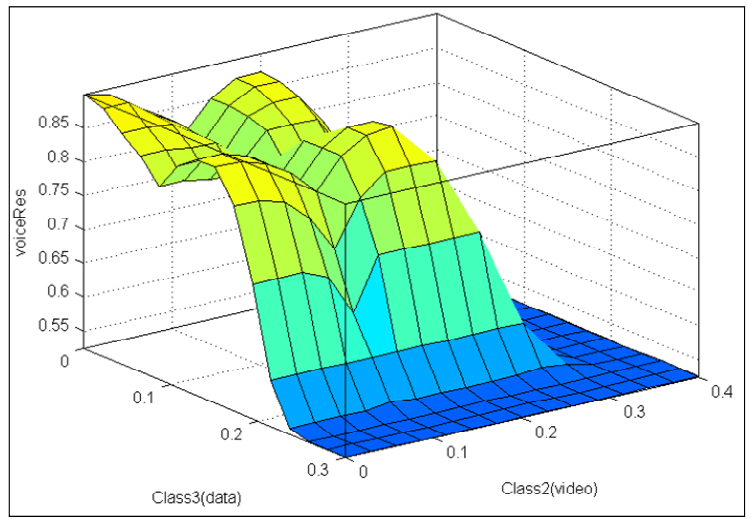

**Figure 11.** Sample output Resource allocation surface map for Class 1 with varying Class 2 and Class 1 network loads.

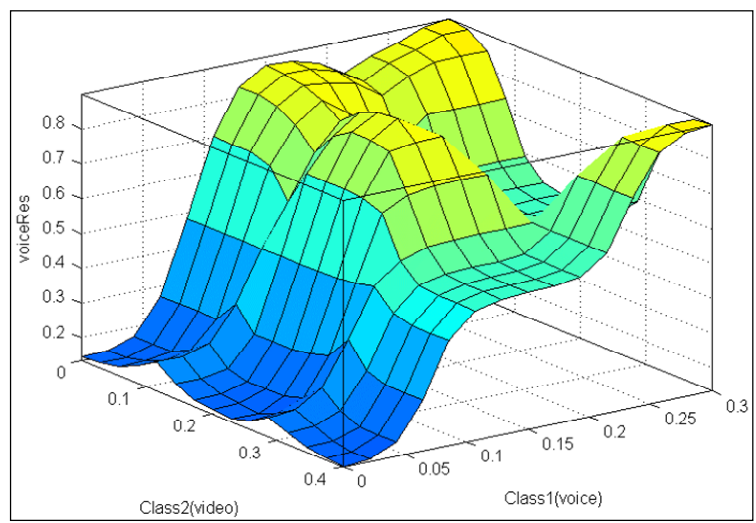

Note that the FRB input membership functions are "scaled" and "shaped" by the policy allocations of $B_1$, $B_2$, and $B_3$ (*i.e.*, the FRB system is dynamically adjusted with policy re-configuration). This design loosely couples the high level resource allocation polices and the fuzzy rule-base system implemented in the RB, so that either one can be independently modified. With three inputs variables each having three membership functions, our current FRB system implementation within CNQF prototype has 27 rules. The $Z^{th}$ rule is expressed as follows:

$R^z$: IF Class 1 load is $F_1^z$ and Class 2 load is $F_2^z$ and Class3 load is $F_3^z$ THEN Class1Res, Class2Res, Class3Res = $y_1$, $y_2$, $y_3$; where $F_1, F_2, F_3$ = {*LL, ML, HL*}, $y_1$, $y_2$, $y_3$ = {*LA, MA, HA*} and Z = 1,…, 27.

*6.3. Experimental Set-up with CNQF Testbed*

This section describes the experiment set up to investigate the efficacy of the FRB system implemented within the CNQF prototype. The experiments use real traffic flows through the testbed shown in Figure 2 and described earlier in Section 4.2. Three classes of traffic flows are generated *i.e.*, voice, video and data and assigned to Classes 1, 2 and 3 respectively. Each voice flow is generated with a bitrate of 32 kbps after an exponentially distributed mean inter-arrival time $1/\lambda_1$ (Poisson arrival rate $\lambda_1$) between flows while the session is terminated after an exponentially distributed service time of $1/\pi_1$ seconds. Class 2 video flows are generated at 384 kbps bitrate with $1/\lambda_2$ exponentially distributed inter-arrival times and exponentially distributed $1/\pi_2$ seconds lifetime, while Class 3 data flows are generated with 256 kbps bitrate with $1/\lambda_3$ exponentially distributed mean inter-arrival time and exponentially distributed mean session time of $1/\pi_3$ seconds.

Aggregate bandwidth measurements per class are pre-processed and periodically input to the FRB system which then computes $B_{FLS}^j$. The bandwidth measurements are sampled every 10 seconds (the minimum possible due to the SNMP protocol operation on the Linux-based testbed). Subsequently, $B_{th}^j$ is determined from Equation (3) to obtain the maximum current allocations per class that are used by the admission control entity to decide whether to accept or reject an arriving session request. Within the high level admission control policies, $B_1$ (for voice traffic) is set at $0.3B_T$, while $B_2$ is set at $0.4B_T$ for video traffic and $B_3 = 0.3B_T$ is assigned to data traffic. $B_T$, the maximum total available bandwidth for all the three classes is set at 6400 kbps for the experiments.

The parameters and rules of the FRB are coded (as described earlier) in an external file which is read by the FRB module in the initial bootstrap phase. Thus, our implementation allows the entire rules and parameters to be re-written and updated offline if desired without disrupting the runtime CNQF admission control functionality. A screen shot of monitoring graphs from the CNQF management user interface depicting instantaneous traffic load per class during the experiments is shown in Figure 12.

*6.4. Experimental Results and Discussions*

For all the experiments undertaken, an identical average inter-arrival time of 1.25 s ($\lambda_1 = \lambda_2 = \lambda_3 = 0.8$) are used for the three traffic profiles while the instantaneous load was varied by using different average service lifetimes such that the ratio $\rho$ proportional to the load given by $\rho_i = (bw_i\lambda_i / B_i\pi_i) | \forall i \in [1,2,3]$ was varied from 0.2, 0.4, 0.6 to 0.8. Recall from the traffic profiles that $bw_1 = 32$ kbps, $bw_2 = 384$ kbps and $bw_3 = 256$ kbps; thus for $\lambda_1 = \lambda_2 = \lambda_3 = 0.8$ and the given $B_1$, $B_2$ and $B_3$ values in the policies the average service times used for each class are given in Table 2. Using these parameters, we investigated *service availability, blocking probability and bandwidth utilization* obtained from admission control utilizing the fuzzy-based resource allocation.

In order to benchmark the performance obtainable from the traffic profiles used in studying the implemented fuzzy-based scheme, we considered a scenario where an arriving flow is granted admission regardless of its class of traffic. Note that since identical average arrival rates are used for the three classes, this scenario is expected to yield maximal total service availability for every $\rho$ considered (since admission is made regardless of class based upper limit); hence appropriate for comparative analysis with the implemented fuzzy-based scheme. For the purpose of comparison, we will call this scenario the "*Class-agnostic*" scenario. Another scenario with admission control based on the fixed $B_i$ thresholds stipulated in the policies (*i.e.*, without fuzzy-rule based resource allocation) is also analyzed. For the purpose of comparison, we will call this scenario the "*Base Policy*" scenario. After 10 experimental runs for each of the three scenarios with each $\rho$ value given in Table 2, the average yielded results for the performance metrics depicted in Figures 13 and 14. Each experimental run had a total of 15000 arrivals request with voice, video and data class traffic each having 5000 generated service requests.

**Figure 12.** A CNQF management UI screenshot showing traffic load for the three classes.

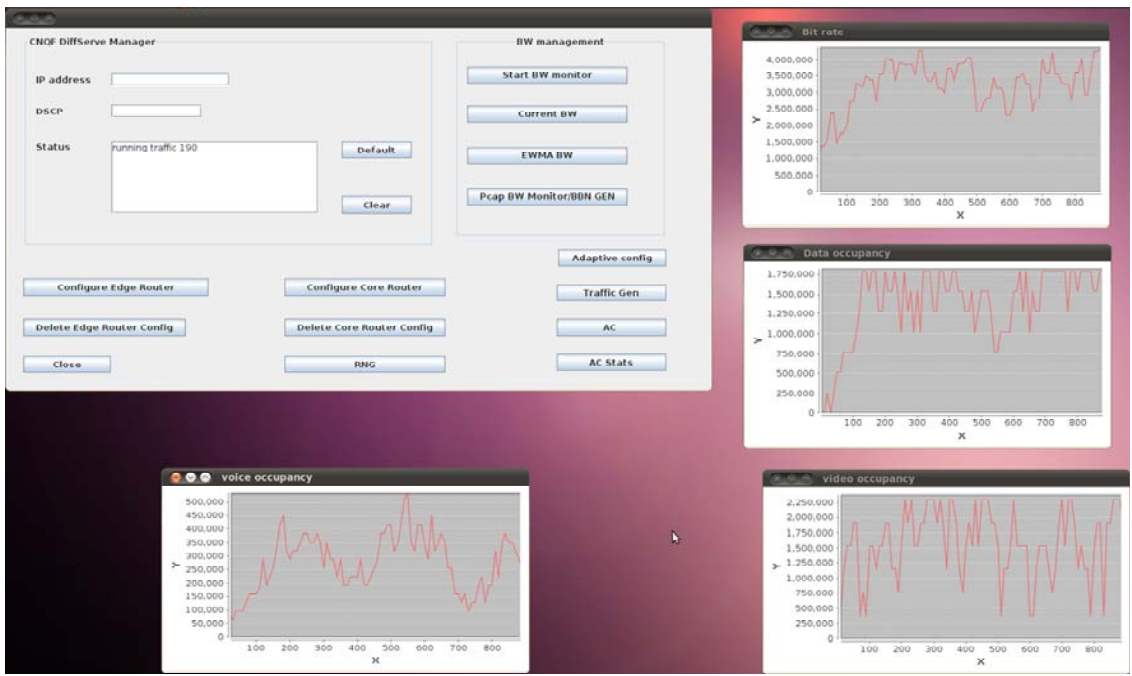

**Table 2.** Average service lifetime values (seconds) per class utilized in the experiments.

| $\rho$ | 0.2 | 0.4 | 0.6 | 0.8 |
|---|---|---|---|---|
| $1/\pi_1$ (s) | 15 | 30 | 45 | 60 |
| $1/\pi_2$ (s) | 1.6665 | 3.333 | 5 | 6.666 |
| $1/\pi_3$ (s) | 1.875 | 3.75 | 5.625 | 7.5 |

From Figure 13, it can be seen that the FRB implementation allows better overall service availability for all the considered $\rho$ representing a range of loading from low to high. This translates to potential higher revenue for an operator utilizing the fuzzy-rule based approach. As expected, service availability drops with higher traffic loading in all three scenarios.

The probabilities of denied service request (blocking probabilities) per class are shown in Figures 14, 15 and 16. That of voice traffic can be seen in Figure 14 where the FRB scheme allows maximum service availability for voice traffic with $\rho = 0.2$ but about only 4% of the voice requests were denied with $\rho = 0.4$ compared to 15% when requests are granted without consideration for class or application of the FRB allocation. The implemented FRB resource allocation scheme is seen to improve the voice service request from Figure 14 over the Class-agnostic admission even at higher traffic loading. The reason why the Base Policy scenario performs more favorably for the voice traffic as ρ is increased is because of the relatively low bandwidth requirement which reduces its overall probability of not obtaining enough bandwidth for its session whilst the more bandwidth hungry video and data class flows have been curtailed by the base policy thresholds. Figures 15 and 16 depict clearly the performance improvement achieved as a result of employing the fuzzy-rule based resource allocation scheme. With reduced blocking probability, more video and data requests can be admitted and potential revenue increase can accrue for service providers whilst maintaining high level of quality of user experience.

**Figure 13.** Overall service availability plot for the three scenarios.

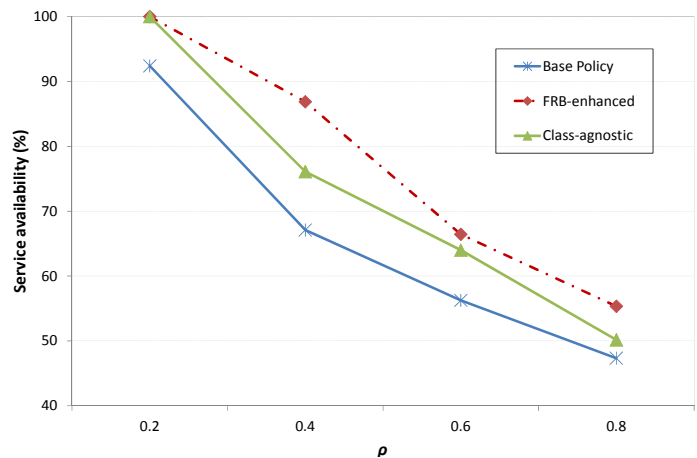

From Figure 17, the cumulative distribution functions of the total bandwidth utilization during a single experimental run with $\rho = 0.2$ for the three scenarios are shown. The plots reveal performance improvement of the FRB system over the other two scenarios in terms of bandwidth utilization. Indeed, the system performance (service availability/blocking probability) can be tuned by changing the rules in the fuzzy system to attain desired performance goals and/or optimize the utilization of one or more of the service classes over others if for example there is a particular incentive in doing so, say cost reduction or revenue improvement.

**Figure 14.** Class 1 (voice) services blocking probabilities for the three scenarios.

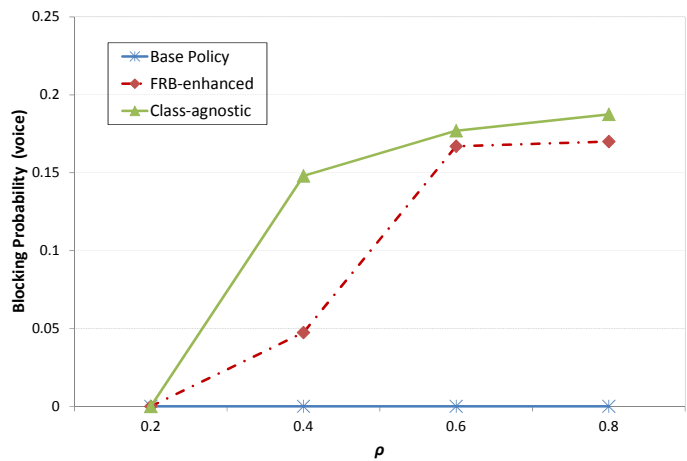

**Figure 15.** Class 2 (video) services blocking probabilities for the three scenarios.

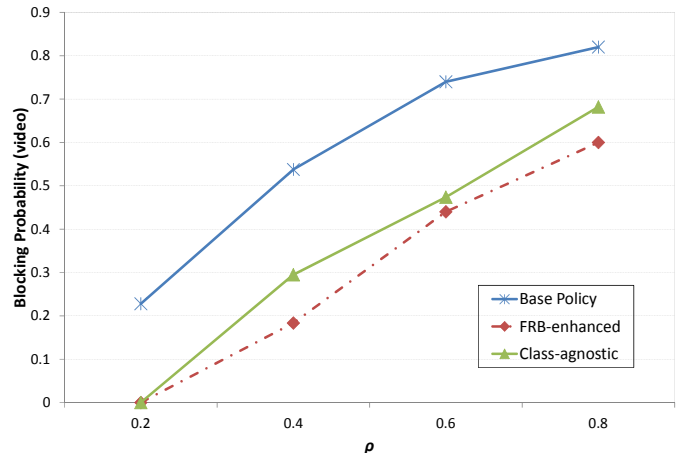

**Figure 16.** Class 3 (data) services blocking probabilities for the three scenarios.

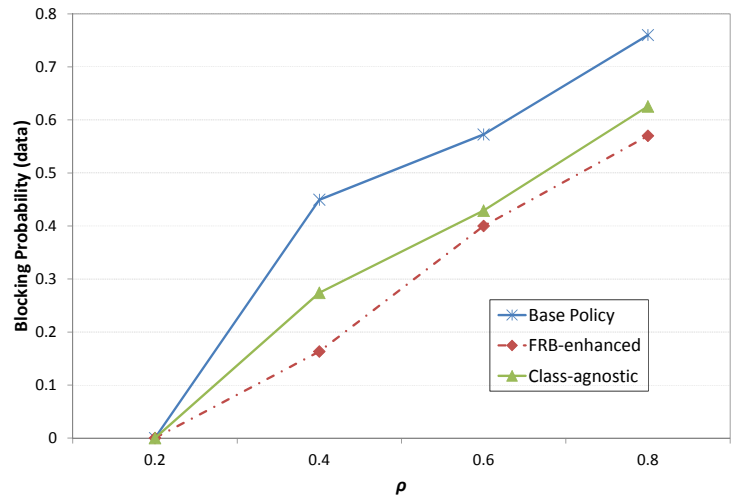

**Figure 17.** Cumulative Distribution Functions of total bandwidth utilization for $\rho = 0.2$.

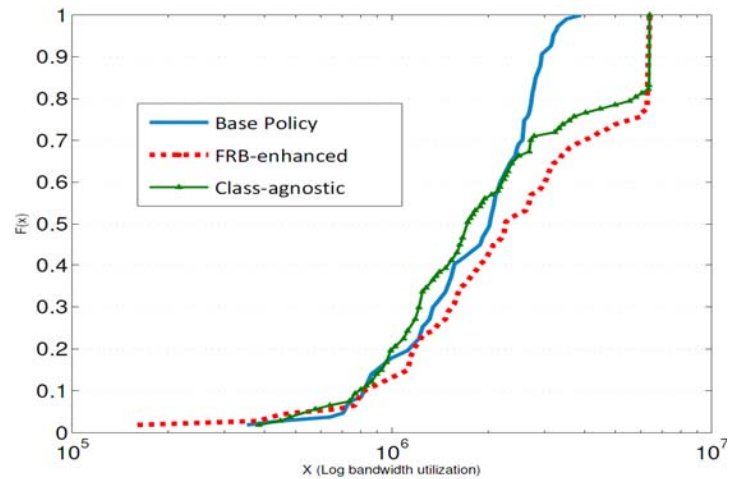

## 7. Conclusions

In this paper we presented a proposed PBNM framework for end-to-end QoS control in converged networks, CNQF (Converged Networks QoS Management Framework), together with its implementation and enhancements which have been evaluated on real testbed environments. CNQF is designed to provide homogenous unified QoS management and resource control over heterogeneous access technologies by leveraging PBNM paradigm which allows transport plane technology-independent solutions to be developed as we have demonstrated in this paper. The potential for PBNM to simplify the administration of complex network characteristics such as QoS, access control, security, resource allocation *etc.* has driven the standardization of architectures within the IETF, 3GPP and TISPAN which all bear similarities. Whilst the standards preclude implementation details and experience of domain specific applications, several studies can be found proposing PBNM in different contexts such as cloud environments, *ad hoc* networks, VPNs, *etc.* Beyond previously presented PBNM studies, in this paper we have proposed means for more intelligent and adaptive management of network resources in cost efficient manner by incorporating fuzzy logic system as a key underpinning enhancement to our CNQF policy based management framework.

By leveraging fuzzy logic there is the potential to simplify network management as well as reduce monitoring load requirements as intelligent solutions can be derived through the use of imprecise data instead of crisp measurements. Furthermore, fuzzy rule based systems provide a convenient and flexible means to couple dynamic operational policies with resource management strategies in an extensible and easily modifiable manner without disrupting normal operation. Our implementation of fuzzy-rule-based driven resource allocation within the CNQF policy based admission control process demonstrated the efficacy of this approach. We further tested the performance of the fuzzy based system implementation compared to scenarios without the approach using real traffic flows on our testbed. The results of the experiments showed that not only is it possible to gain improved service availability and lower the blocking probability for high revenue bandwidth hungry services such as video, but also improved bandwidth resource utilization can also be obtained from the approach. Nevertheless, the flexibility in tuning the resource allocation without ongoing service disruption via fuzzy rule base and the allocation policies to potentially achieve even better performance is a promising prospect. Service classes can also be easily extended leading to a more scalable and cost-effective resource management solution for PBNM systems which are likely to be a dominant network management paradigm in next generation converged networks. In our future work, we would like to extend the fuzzy rule based system to support and enhance context-aware resource allocation and QoS control in the PBNM system and also evaluate the performance.

**Acknowledgement**

This work is funded by the EPSRC-DST India-UK Advanced Technology Centre of Excellence in Next Generation Networks, Systems and Services (IU-ATC) (www.iu-atc.com) under grant EP/G051674/1.